\documentclass[aps,prb,twocolumn,superscriptaddress]{revtex4-2}
\usepackage[utf8]{inputenc}
\usepackage{amsmath}
\usepackage{graphicx}
\usepackage{subfigure}
\usepackage{xcolor}
\usepackage{amsfonts}
\usepackage{booktabs}
\usepackage{bigstrut,multirow,rotating}
\usepackage{appendix}
\usepackage{svg}
\usepackage{hyperref}
\hypersetup{
  colorlinks=true,
  citecolor=magenta,
  linkcolor=red,
  urlcolor=cyan}

\def\Jth{J_{\frac{3}{2}}}
\def\Qth{Q_{\frac{3}{2}}}
\def\Jf{J_{{\rm SU}(4)}}
\def\Qf{Q_{{\rm SU}(4)}}

\begin{document}

\title{First-order N\'eel-VBS transition in $S=3/2$ antiferromagnets}

\author{Fan Zhang}
\affiliation{Department of Physics, Beijing Normal University, Beijing 100875, China}

\author{Wenan Guo}
\email{waguo@bnu.edu.cn}
\affiliation{Department of Physics, Beijing Normal University, Beijing 100875, China}
\affiliation{Key Laboratory of Multiscale Spin Physics (Ministry of Education), Beijing Normal University, Beijing 100875, China}
\author{Ribhu K. Kaul}
\email{ribhu.kaul@psu.edu}
\affiliation{Department of Physics, The Pennsylvania State University, University Park PA-16802, USA}

\date{\today}

\begin{abstract}
We study the transition between N\'eel and columnar valence-bond solid ordering in two-dimensional $S=3/2$ square lattice quantum antiferromagnets with SO(3) symmetry. According to the deconfined criticality scenario, this transition can be direct and continuous like the well-studied $S=1/2$ case. To study the global phase diagram, we work with four multi-spin couplings with full rotational symmetry, that are free of the sign-problem of quantum Monte Carlo. Exploring the phase diagram with quantum Monte Carlo simulations, we find that the phase transition between N\'eel and valence-bond solid is strongly first-order in the parts of the phase diagram that we have accessed.      
\end{abstract}

\maketitle

\section{Introduction}

The effect of the spin quantum number $S$ on the phase diagram of spin models was brought to focus in Haldane's seminal work on one-dimensional spin chains~\cite{Haldane1983, Haldane1983-2}. The parameter $S$ appears as a co-efficient of a topological term in the effective field theory of spin chains and simply changing it from half-odd integer to integer has dramatic effects on the physical properties, for example changing the spectra from gapless to gapped (for a pedagogical introduction, see~\cite{sachdev:qpt}). This even-odd dependence on $S$ in one-dimension has been well studied and is firmly established in various microscopic models of spin chains, see e.g.~\cite{HWJ1986-spin1chain, AKLT1988, affleck1986-Haldane, Huse1993-spin1chain, kennedy1990, Lacaze1994}.

On two-dimensional (2D) square lattices, the spin-$S$ also plays an important role in the field theory. Haldane showed that half-odd integer spins result in a quadrupling of hedgehogs in the effective field theory, odd integer spins cause a doubling of hedgehogs and even integer spins do not affect the action of hedgehogs~\cite{haldane1988}. While these differences are not expected to affect the ordered N\'eel state, they can have dramatic consequences on the quantum disordered phases and the phase transitions between them. One consequence of this difference was put forth as the ``deconfined criticality scenario" which predicted direct continuous transitions between N\'eel and four-fold valence-bond solid(VBS) ordered states in $S=1/2$ square lattice anti-ferromagnets~\cite{senthil2004-DQCP}. Since the original proposal, there has been an intense study of its validity using large-scale numerical simulations of $S=1/2$ antiferromagnets (see e.g.~\cite{Sandvik2007-JQ, melkokaul, nahumso5, 
demidiofirst, takahashi2024so5multicriticalitytwodimensionalquantum}). What is the effect of the higher spin-$S$ on the $S=1/2$ deconfined scenario?
In comparison to $S=1/2$, work on the extensions to higher-$S$ in two dimensions is limited. An extension of the theory to $S=1$ square lattice model studied the phase transition between N\'eel and a nematic state, with a possible critical point~\cite{wangfa_NP}. Sign free $S=1$ microscopic models for the transitions from N\'eel to VBS~\cite{julia2020}, and N\'eel to the nematic state~\cite{Nisheeta} have been studied and the transitions found to be first order. Moving to $S=3/2$, according to Haldane's Berry phase evaluation all half-odd integer spin antiferromagnets should be described by the same universal field theory, hence it is expected from field theory that $S=3/2$ antiferromagnetic models should exhibit deconfined criticality very similiar to the $S=1/2$ models, although this expectation has not been studied numerically.

With this motivation, our purpose in this paper is to study the N\'eel-VBS transition in square lattice $S=3/2$ models using Monte Carlo methods. There has been very limited numerical work on microscopic models of $S=3/2$ in two dimensions -- the larger Hilbert space is inconvenient for numerical methods based on diagonalization, and the space of sign problem-free models and the Monte Carlo algorithms to simulate them is less well-developed than for $S=1/2$. Here, we use the split-spin representation~\cite{todokato} and its use to design sign-free spin-$S$ models~\cite{Nisheeta} to construct efficient simulation algorithms for microscopic models of the phase transition.

This paper is structured as follows: In Section \ref{sec:model}, we introduce the models studied. We then delve into the split-spin basis of the Stochastic Series Expansion (SSE) Quantum Monte Carlo (QMC) algorithm and elucidate the process of rewriting the Hamiltonian on this basis.
Following this, Section \ref{sec:order} studies the finite-size scaling behavior of both N\'eel and VBS order parameters. Remarkably, the findings from Section \ref{sec:order} provide compelling evidence of a direct first-order N\'eel-VBS phase transition. Finally, in Section \ref{sec:sum}, we provide a comprehensive summary and offer insights into potential future research directions.

\section{Model and method\label{sec:model}}

\begin{figure}[!t]
\centering
\includegraphics[width=0.45\textwidth]{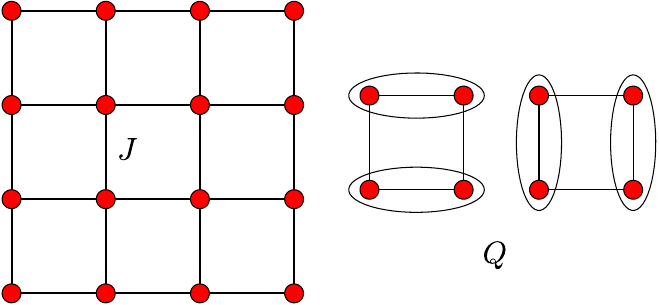}
\caption{\label{lattice}Illustrating the lattice representation of the $J$-$Q$ model. ``$J$" denotes the antiferromagnetic interactions between adjacent sites, and ``$Q$" denotes the product of two singlet projection operators acting on four neighboring sites.}
\end{figure}

The spin-1/2 $J$-$Q$ model is written as\cite{Sandvik2007-JQ}:
\begin{equation} \label{eq:H_JQ}
H=J\sum_{\langle ij\rangle} \vec S_i\cdot \vec S_j-Q \sum_{\langle ijkl \rangle} \left( \frac 14 -\vec S_i\cdot \vec S_j\right)\left( \frac 14 -\vec S_k\cdot \vec S_l\right),
\end{equation}
where $\vec S_i$ refers to a $S=1/2$ spin at site $i$ in a 2D square lattice. $\left( \frac 14 -\vec S_i\cdot \vec S_j\right)$ is a spin singlet projection operator acting on two neighboring $S=1/2$ spins. It eliminates the triplet states and keeps the singlet state when acting on the two spins. 
A quantum phase transition from the N\'eel state to the columnar VBS is realized by tuning the ratio $Q/J$. 
The summation $\langle ij \rangle$ denotes all nearest two neighbors $i,j$ on a bond, as depicted by $J$ in Fig. \ref{lattice}. Similarly, the summation $\langle ijkl \rangle$ represents all nearest four neighbors $i,j,k,l$ within a plaquette, illustrated by $Q$ in Fig. \ref{lattice}. 

The naive generalization of the spin-1/2 $J$-$Q$ model to the $S=3/2$ Hilbert space is the $\Jth$-$\Qth$ model:
\begin{equation}\label{eq:H_JQJ}
H_{\frac{3}{2}}=\Jth\sum_{\langle ij\rangle} \vec S_i\cdot \vec S_j-\Qth \sum_{\langle ijkl \rangle} \left( \frac 94 -\vec S_i\cdot \vec S_j\right)\left( \frac 94 -\vec S_k\cdot \vec S_l\right),
\end{equation}
with the $\vec S_i$ are now $S=3/2$ operators. The only other difference from the $S=1/2$ model is the factor of $9/4$. Now however, that $\left( \frac 94 -\vec S_i\cdot \vec S_j\right)$ is not a singlet projector. When acting on two neighboring spin-3/2 spins, it preserves all states except for the total $S=3$ state. The disparity between two $S=1/2$ spins and two $S=3/2$ spins lies in their respective states: the former encompasses singlet and triplet states, while the latter introduces a variety of additional states. In any case, numerically, we find that the $\Jth$-$\Qth$ model fails to realize a phase transition from N\'eel to VBS being N\'eel ordered for all couplings. The numerical evidence is presented in Appendix \ref{JQJmodel}. 

As noted above, the Heisenberg interaction is not a singlet projector for $S=3/2$. Indeed the two-site singlet projector is a new  SU(2) invariant interaction,  
 $P(\vec S_i +\vec S_j)$. Its explicit form is,
\begin{equation}\label{eq:P}
P(\vec S)=-\frac{(S^2-12)(S^2-6)(S^2-2)}{144}.
\end{equation}
The operator $P(\vec S_i +\vec S_j)$, when applied to a system composed of two spin-3/2 sites, results in the elimination of all states except for the singlet state:
\begin{equation}
\chi_{0,0}=\frac 12 \left( \left | -\frac 32 , \frac 32 \right\rangle - \left | -\frac 12 , \frac 12 \right\rangle + \left | \frac 12 , -\frac 12 \right\rangle - \left | \frac 32 , -\frac 32 \right\rangle \right).
\end{equation}
This operator clearly has SU(2) invariance. Indeed, it has a larger SU(4) symmetry (in the staggered fundamental-conjugate to fundamental representation) of which the SU(2) is a subgroup.

Using this operator, we can introduce two more couplings that act on our $S=3/2$ Hilbert space, $\Jf$ and $\Qf$,
\begin{eqnarray}
H_{{\rm SU}(4)}&=&-\Jf\sum_{\langle ij\rangle}  P(\vec S_i+\vec S_j) \nonumber \\
&-&\Qf \sum_{\langle ijkl \rangle} P(\vec S_i+\vec S_j)P(\vec S_k+\vec S_l),
\end{eqnarray}
The $\Jf$-$\Qf$ model has been studied previously and hosts a N\'eel-VBS transition in the context of SU(4) deconfined criticality~\cite{Lou-SUN,kaulsu34}. In this work, we will focus instead on the SU(2) $S=3/2$ criticality, which requires us to have some of the terms in $H_{\frac{3}{2}}$ finite to lower the symmetry from SU(4) to SU(2).

In the large space of four couplings we have introduced above ($\Jth,\Qth,\Jf,\Qf$), we will focus here on the phase diagram of the $\Jth$-$\Qf$ model, which is tuned by one parameter $g\equiv \Jth/\Qf$. We chose to work with these coupling because we know, at $g=\infty$, the system must be in the N\'eel state and the $g=0$ state is in the VBS~\cite{Lou-SUN,kaulsu34}. At any finite value of $g$ this model has only the SU(2) symmetry with a four-dimensional Hilbert space appropriate to $S=3/2$, and is hence an appropriate model for SU(2) deconfined criticality. We have studied other combinations of couplings and included the results for completeness in Apps.~\ref{JQJmodel}, \ref{JQQJmodel}, \ref{QJQmodel}.

We employ the QMC method based on the SSE representation \cite{Sandvik-SSE} to simulate our system. For the $J$-$Q$ model with $S>1/2$, the Directed Loop method \cite{Sandvik-DL1999, Sandvik-DL2002} is typically required.
In this paper, we use the split-spin\cite{todokato} method to simulate the spin-3/2 $\Jth$-$\Qf$ model. We rewrite the spin-3/2 on each of the $N$ lattice sites as three spin-1/2 ``mini-spins",
\begin{equation}
\vec S_i=\sum_{a=1}^3 {\vec s}_i^{a}.
\end{equation}
Here $\vec s_i^a$ has a lattice index $i$ with $1 \leq i \leq N$ and a minispin index $a$ with $1\leq a \leq 3$.
The $N$ $S=3/2$ spins then become $3N$ $S=1/2$ spins. The dimension of the Hilbert space also changes from $4^N$ to $2^{3N}$. To faithfully simulate the original problem, we have to include a projection operator, $\mathcal{P}=\prod_i \mathcal{P}_i$, where $\mathcal{P}_i$ projects out the spin-3/2 from the mini-spin basis. This method makes the simulation of the models relatively simple.

Using the mini-spin basis, the $\Jth$ term can be expressed as:
\begin{equation}
H_{ij}^{\Jth}=\vec S_i \cdot \vec S_j=-\sum_{a,b} \left(\frac 14 -\vec s_i^a\cdot \vec s_j^b \right).
\end{equation}
It can also be proven that the singlet projector on two $S=3/2$ spins can be expressed as:
\begin{equation}
\begin{split}
P&(\vec S_i+\vec S_j) = P_{\frac 32} \left( \sum_{a=1}^3 \vec s_i^a\right ) P_{\frac 32} \left( \sum_{b=1}^3 \vec s_j^b\right ) \times\\
&\frac {1}{18}\sum_{\substack{a\ne c\ne e \\ b\ne d\ne f}} \left(\frac 14 -\vec s_i^a\cdot \vec s_j^b \right) \left(\frac 14 -\vec s_i^c\cdot \vec s_j^d \right) \left(\frac 14 -\vec s_i^e\cdot \vec s_j^f \right),
\end{split}
\end{equation}
where $P_{\frac 32} \left( \sum_{a=1}^3 \vec s_i^a\right )$ projects out the spin-3/2 from the minispin basis.
It is worth noting that all interactions occur between mini-spins of different sites $i$ and $j$.
Details of this algorithm can be found in \cite{Nisheeta}.

In this work, we define $g=\Jth/\Qf$ and set  $\Jth^2+\Qf^2=1$,
choosing $\beta=L$ in the simulations to study the quantum phase transition.

\section{Results\label{sec:order}}

The model is expected to exhibit a N\'eel phase for $g$  large enough and a VBS phase when $g$ is small. By measuring quantities sensitive to the N\'eel and VBS orders, we confirm the existence of these two phases and investigate the phase transition.
We find that the Binder cumulants of these two order parameters approach 1 in the ordered phase and 0 in the disordered phase, and exhibit non-analytic and negative divergence as the system size approaches infinity. We also examine the properties of the phase transitions from the histograms of the order parameters. All these results are consistent with a first-order phase transition.

\subsection{\label{op}Order Parameters}

In this section, we study the finite-size scaling of the N\'eel and VBS order parameters. We are limited to somewhat small system sizes $L\leq 20$ close to the transition because of metastability issues associated with a first-order transition. Metastability is interesting in itself and explored in Sec.~\ref{metast}.

\begin{figure}[t]
\centering
\setlength{\abovecaptionskip}{20pt}
\includegraphics[angle=0,width=0.5\textwidth]{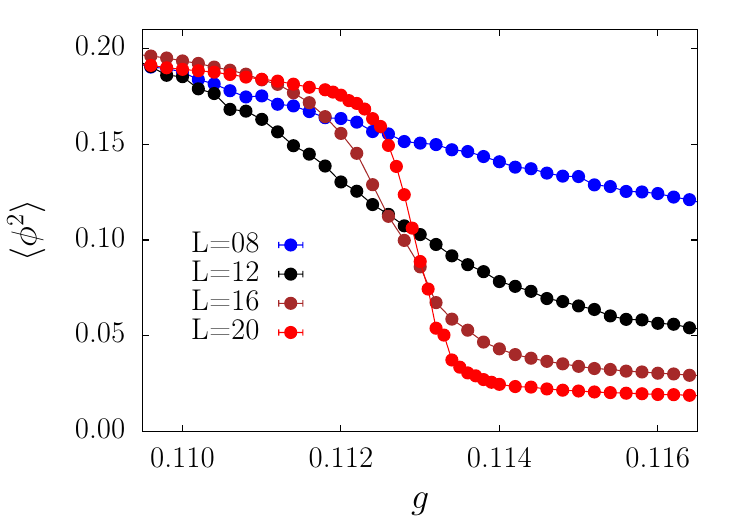}
\caption{The square VBS order parameter $\langle \phi^2 \rangle$ for different system sizes  versus the tuning parameter $g$. A region in which $\langle \phi^2 \rangle$ is roughly volume dependent is separated from a region where $\langle \phi^2 \rangle$ appears to vanish, indicating a phase transition around $g\approx 0.113$. The location and nature of the VBS transition are explored further in Fig.~\ref{uphi} using the Binder cumulant. }
\label{phi2}
\end{figure}

The VBS phase, which breaks the $Z_4$ symmetry, can be characterized by $\langle \phi^2 \rangle$, with the VBS order parameter $\vec \phi=(\phi_x, \phi_y)$ defined as:
\begin{equation}\label{eq:phix}
\phi_x=\frac 1N \sum_{\vec r} S_{\vec r}^z S_{\vec r+\hat x}^z e^{-i\vec k\cdot \vec r}
\end{equation}
and
\begin{equation}\label{eq:phiy}
\phi_y=\frac 1N \sum_{\vec r} S_{\vec r}^z S_{\vec r+\hat y}^z  e^{-i\vec k\cdot \vec r}.
\end{equation}
The wave vectors are $\vec k = (\pi,0)$ for $\phi_x$ and $(0,\pi)$ for $\phi_y$. $\hat{x}$ and $\hat{y}$ represent neighboring sites in the $x$ and $y$ directions, respectively.  Figure \ref{phi2} depicts $\langle \phi^2 \rangle$ in the vicinity of the phase transition point for various system sizes. The decrease of $\langle \phi^2 \rangle$ from a finite value to 0 as $g$ increases suggests the presence of a phase transition from a VBS ordered to a VBS disordered phase around $g\approx0.113$.

\begin{figure}[t]
\centering
\setlength{\abovecaptionskip}{20pt}
\includegraphics[angle=0,width=0.5\textwidth]{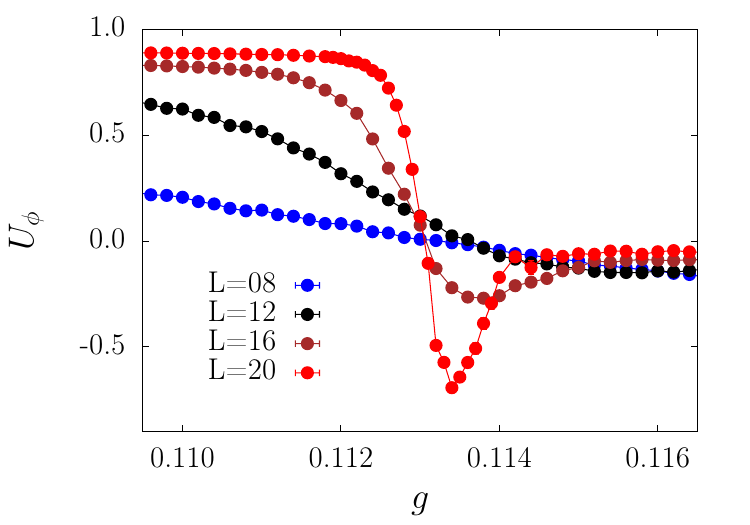}
\caption{The Binder cumulant $U_\phi$ for different system sizes as functions of $g$. $U_\phi$ tends to 1 in the ordered phase and 0 in the disordered phase, indicating a phase transition at $g\approx0.113$. $U_\phi(L)$ exhibits a pronounced negative trend in proximity to the phase transition point characteristic of a first-order phase transition.}
\label{uphi}
\end{figure}

We define the Binder cumulant $U_\phi$\cite{Binder1981, Binder1981-2}, which serves as a useful quantity to study the transition, in particular, distinguishing between continuous and first-order phase transitions,
\begin{equation}
U_{\phi}=2-\frac{\langle {\phi}^4\rangle}{\langle {\phi}^2\rangle^2}. 
\end{equation}
$U_{\phi}$ tends towards 1 in the ordered phase and towards 0 in the disordered phase as $L$ approaches infinity. For a continuous phase transition, for sufficiently large $L$, $U_\phi$ varies continuously between 0 and 1 with the tuning parameter and converges to a fixed point value for different system sizes at the transition point, while for a first-order transition, $U_\phi$ exhibits non-analyticity and negativity, tending towards negative infinity near the phase transition point as size increases to infinity.  
Figure \ref{uphi} displays $U_\phi$ for various system sizes in the vicinity of the phase transition point, showing the characteristics of a first-order phase transition\cite{Binder-firstorder}.

\begin{figure}[htbp]
\centering
\setlength{\abovecaptionskip}{20pt}
\includegraphics[angle=0,width=0.5\textwidth]{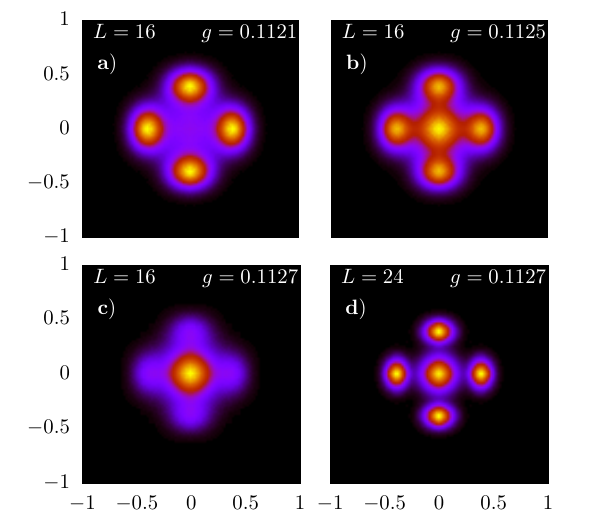}
\caption{The histograms of $\vec\phi$. a), b), and c) for $L=16$ at VBS phase ($g=0.1126$), at around phase transition point ($g=0.1123$), and at VBS phase ($g=0.1121$), respectively; d) for $L=24$, at around phase transition point ($g=0.1127$).}
\label{hisphi}
\end{figure}

Since at the transition point of a first-order phase transition two distinct phases coexist, a direct method of detecting a first-order phase transition is sampling the histograms of order parameters. The distribution function $P(\phi_x, \phi_y)$ should exhibit peaks at $(0,0)$ in the VBS disordered phase, and at $(\pm \phi_0, 0)$ and $(0, \pm \phi_0)$ in the VBS ordered phase due to the $Z_4$ symmetry, where $\phi_0$ represents a finite value. Figure \ref{hisphi} illustrates the histograms of $\vec{\phi}$ for $L=16$ in the VBS phase, around the phase transition point, and in the VBS disordered phase, respectively; and the histogram for $L=24$ around the phase transition point. 
The coexistence of peaks at $(0,0)$ and $(\pm \phi_0, 0)$, $(0, \pm \phi_0)$, and the sharpening of the peaks with increasing system sizes provide further evidence of a first-order phase transition from the perspective of the VBS order parameter.

\begin{figure}[!b]
\centering
\setlength{\abovecaptionskip}{20pt}
\includegraphics[angle=0,width=0.5\textwidth]{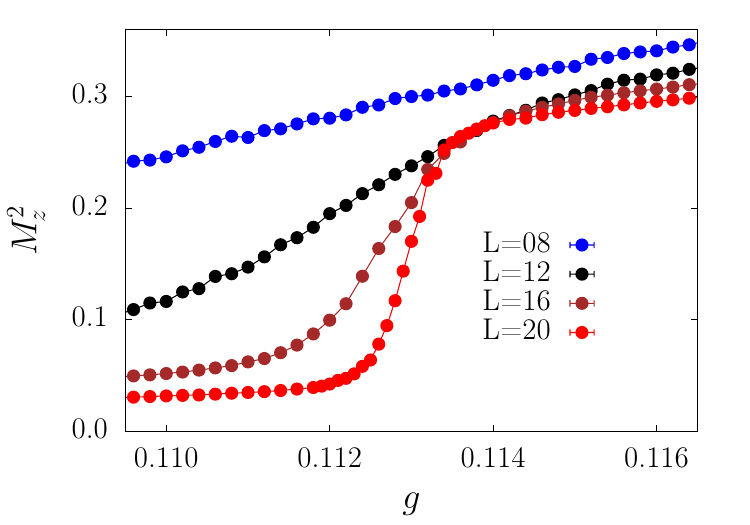}
\caption{The squared staggered magnetization $M_z^2$
for different system sizes versus $g$, showing two regions. For $g$ smaller than approximately 0.113, the order parameter scales to zero, and for $g$ larger than this value, it appears to scale to a finite value. The transition point for the N\'eel order parameter is analyzed further in Fig.~\ref{um}. }
\label{msz2}
\end{figure}

 We now turn to the behavior of the N\'eel order parameter close to the transition. The magnetically ordered phase with the $O(3)$ spin rotational symmetry broken can be characterized by the N\'eel order parameter $m_s^z$, which is the $z$ component of staggered magnetization of the system.
\begin{equation}\label{eq:msz2}
m_s^z=\frac 1N \sum_{\vec r} S^z_{\vec r} e^{-i\vec k \cdot \vec r} 
\end{equation}
with $\vec k=(\pi,\pi)$ is the wave vector corresponding to the N\'eel phase, $N=L^2$. This quantity is diagonal in the $S^z$ basis and easy to measure in the SSE simulations. For finite-size systems, 
the square staggered magnetization $M_z^2=\langle (m_s^z)^2\rangle$ is calculated to describe the order.  
Figure \ref{msz2} illustrates $M_z^2$ 
in the vicinity of the phase transition point for various system sizes. The increase of $M_z^2$ from 0 to a finite value as $g$ increases suggests the presence of a phase transition from an N\'eel disordered to a N\'eel ordered phase around $g=0.113$.

\begin{figure}[!t]
\centering
\setlength{\abovecaptionskip}{20pt}
\includegraphics[angle=0,width=0.5\textwidth]{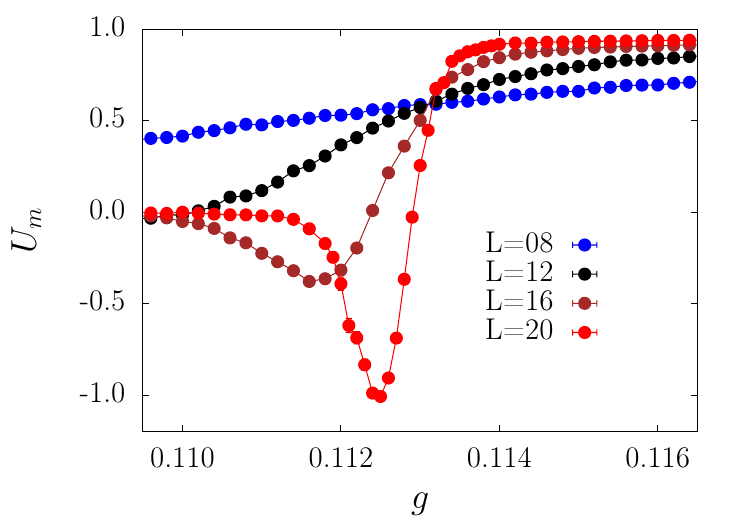}
\caption{The Binder cumulant of the staggered magnetization $U_m$ for different system sizes as functions of $g$. The curves exhibit a pronounced negative trend in proximity to the transition point, indicating a first-order transition. The transition point for the N\'eel and order is consistent with the transition point for the VBS order, indicating a direct transition, which is confirmed in the analysis shown in Fig.~\ref{gc}. }
\label{um}
\end{figure}

The Binder cumulant of N\'eel order parameter is defined as:
\begin{equation}
U_m=\frac 56 \left(3-\frac{\langle {(m_s^z)}^4\rangle}{\langle {(m_s^z)}^2\rangle^2} \right).
\end{equation}
Similarly, as depicted in Fig.~\ref{um}, $U_m$ becomes non-analytic, tending towards negative infinity near the transition point, as $L$ increases, indicative of a first-order phase transition.\cite{Binder-firstorder}

\begin{figure}[t]
\centering
\setlength{\abovecaptionskip}{20pt}
\includegraphics[angle=0,width=0.45\textwidth]{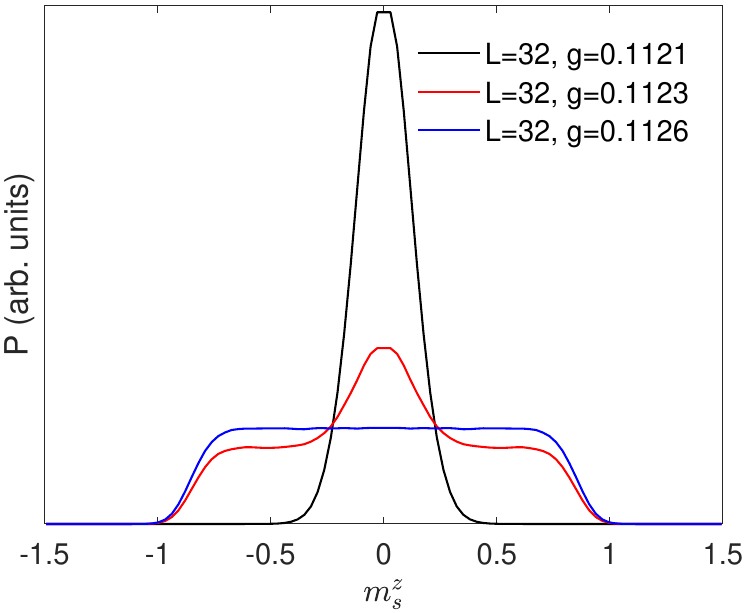}
\vskip 0.4cm
\caption{The histograms of $m_s^z$ for $L=32$ at the N\'eel phase (g=0.1126), around phase transition point (g=0.1123), and the VBS phase (g=0.1121). In the disordered phase a Gaussian distribution is expected, and in the ordered phase a flat distribution. The distribution close to the phase transition shows a superposition of these distributions pointing to the coexistence between a N\'eel ordered and a disordered phase. }
\label{hismsz}
\end{figure}

Just as for the VBS order parameter, we have sampled the histogram of the N\'eel order parameter $m_s^z$. The histogram of $m_s^z$ exhibits a Gaussian distribution centered at zero in the N\'eel disordered phase, a uniform distribution in the N\'eel ordered phase (this is because we are measuring one component of a vector order parameter), and a mixture of these two distributions around the phase transition point due to the coexistence of two different phases characteristic of a first-order phase transition. Figure \ref{hismsz} displays the histogram of the $S=3/2$ $\Jth$-$\Qf$ model for the size $L=32$. The observed phenomenon of phase coexistence at $g=0.1123$ serves as compelling evidence for a first-order phase transition.

The crossings of the Binder cumulants should converge to the transition point as system sizes increase. However, since the transition is first-order,
determining the phase transition point from the crossings in this model proves challenging due to their negative divergent behavior near the phase transition point, as illustrated in Fig.\ref{um} and Fig.\ref{uphi}. It is useful to define the order parameter ratios $R_m$ and $R_\phi$, which goes to 1 in the ordered phase and 0 in the disordered phase as the system size $L$ tends to infinity 
but avoids the negativity near the phase transition point. 

The N\'eel ratio averages over $x$ and $y$ N\'eel ratios
\begin{equation}
R_m=\frac 12 \left( R^x_{m} + R^y_{m} \right)
\end{equation}
with 
\begin{equation}
R^a_{m}=1-\frac{C_{m}((\pi,\pi)+{2\pi \over L} {\hat a})}{C_{m}(\pi,\pi) },
\end{equation}
and $a$ labels $x$ and $y$.
Here $C_m(\vec k)$ is the N\'eel structure factor
\begin{equation}
C_{m}(\vec k)=\frac 1 N \sum_{\vec r} \langle S^z_{0} S^z_{\vec r} \rangle e^{-i\vec k \cdot \vec r}, 
\end{equation}
and $\hat a$  correspond to the unit vector in the $x, y$ direction, respectively. 

The VBS ratio average over $x$ and $y$ VBS ratios
\begin{equation}
R_\phi=\frac 12 \left( R^x_{\phi} + R^y_{\phi} \right)
\end{equation}
with 
\begin{equation}
R^x_{\phi}=1-\frac{C^x_{\phi}(\pi,\frac{2\pi}{L}) }{C^x_{\phi}(\pi,0) },
\end{equation}
and 
\begin{equation}
R^y_{\phi}=1-\frac{C^y_{\phi}(\frac{2\pi}{L},\pi) }{C^y_{\phi}(0, \pi) }.
\end{equation}
Here $C^a_\phi(\vec k)$ is the VBS structure factor defined basing on $a$ direction dimer-dimer correlator
\begin{equation}
C^a_{\phi}(\vec k)=\frac 1 N \sum_{\vec r} \langle S^z_{0} S^z_{\hat a} S^z_{\vec r} S^z_{\vec r+\hat a} \rangle  e^{-i\vec k \cdot \vec r},  
\end{equation}
$a=x, y$ correspond to the $x,y$ directions in 2D square lattice.

\begin{figure}[t]
\centering
\setlength{\abovecaptionskip}{20pt}
\includegraphics[angle=0,width=0.5\textwidth]{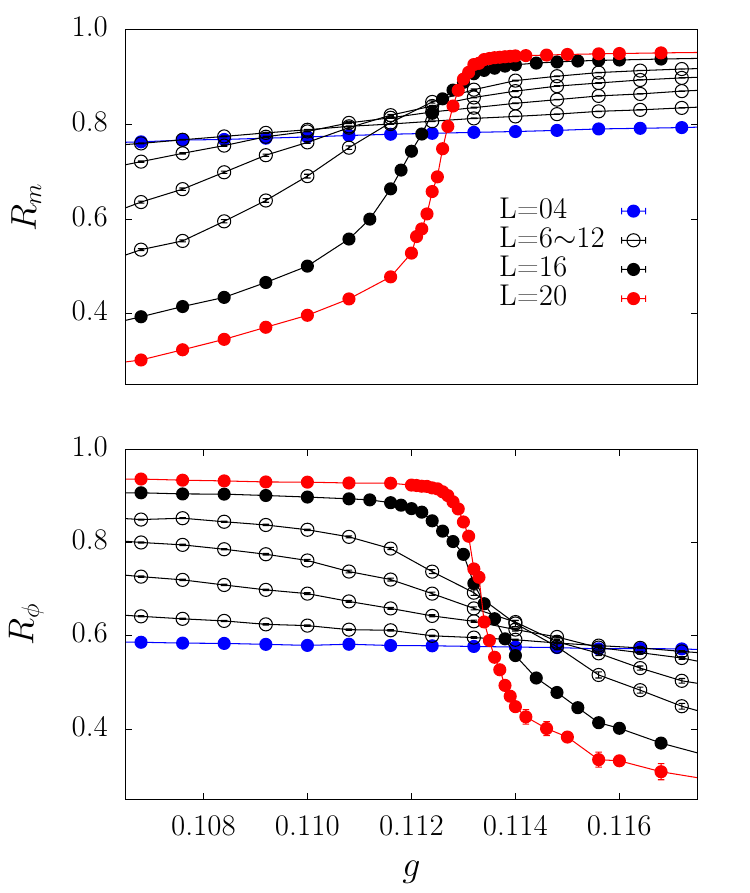}
\vskip 0.4cm
\caption{The order parameter ratios $R_m$ and $R_\phi$ for different system sizes as functions of $g$. }
\label{Rm}
\end{figure}

Figure \ref{Rm} illustrates the order parameter ratios $R_m$ and $R_\phi$ varying with $g$. 
Ensuring the continuity of $R_m$ and $R_\phi$ remains a challenge, particularly as $L$ increases, notably beyond $L>20$, due to long tunneling times between the two phases of the first-order phase transition. Consequently, we opt to utilize smaller system sizes to ascertain the crossing points reliably.

\begin{figure}[!t]
\centering
\setlength{\abovecaptionskip}{20pt}
\includegraphics[angle=0,width=0.5\textwidth]{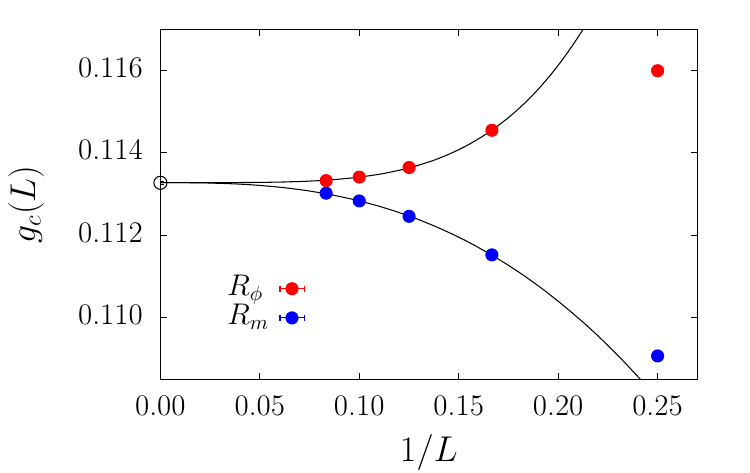}
\vskip 0.4cm
\caption{The crossing points of $(L,2L)$ for $R_m$ and $R_\phi$ with $L=4, 6, 8, 10, 12$. The black solid lines are simultaneously fitted by power laws $g_c(L)^{R_m}=a+b_1/L^{c_1}$ and $g_c(L)^{R_\phi}=a+b_2/L^{c_2}$, within the range of $L=6$ to $12$. This yields $g_c(L\to \infty)=a=0.11327(3)$. The fit has reduced $\chi^2=0.94$.}
\label{gc}
\end{figure}

\begin{figure}[!h]
\centering
\setlength{\abovecaptionskip}{20pt}
\includegraphics[angle=0,width=0.45\textwidth]{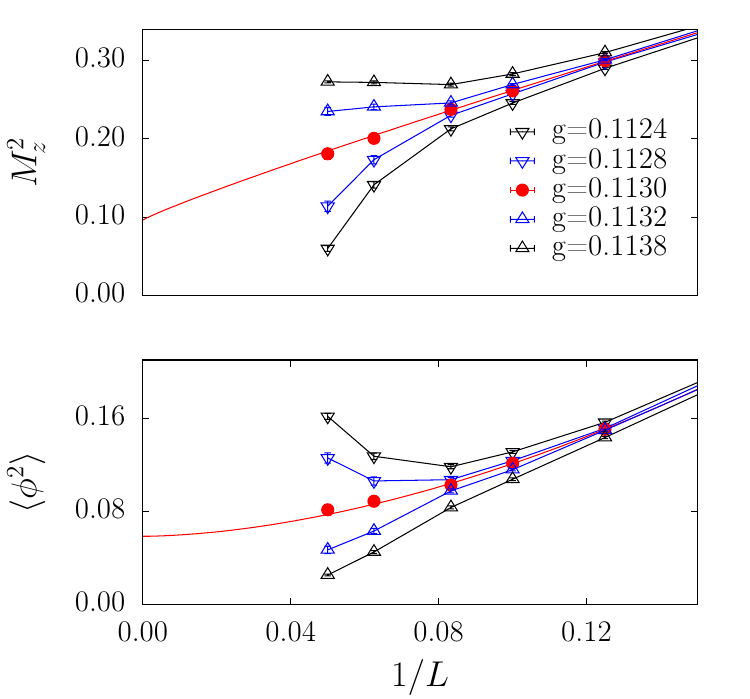}
\vskip 0.4cm
\caption{Finite-size behavior of the order parameters $M_z^2$ and $\langle \phi^2 \rangle$ 
close to the phase transition point up to $L=20$. Power-law extrapolation for both the VBS and the N\'eel order parameters in the form $f(L)=a+b/L^c$ converge to finite values  
at a common coupling strength of $g=0.1130$. The red solid lines represent the fitted functions. 
The fitting results suggest that both order parameters remain finite at the transition point as the system size approaches infinity, indicative of a first-order phase transition, although the quantitative reliability of the fit values may be limited. }
\label{mszl-2}
\end{figure}

Figure \ref{gc} illustrates the crossing points of $(L,2L)$ for $R_m$ and $R_\phi$, where $L$ ranges from $4$ to $12$. These crossings demonstrate a convergence towards a common value when $L\to \infty$, indicating the phase transition point $g_c$. Remarkably, this convergence suggests that 
the phase transition described by two distinct order parameters, $\langle (m_s^z)^2 \rangle$ and $\langle \phi^2 \rangle$, is a single phase transition from a N\'eel-ordered state to a VBS-ordered state. 

The preceding analysis demonstrates that both order parameters exhibit characteristics of a first-order phase transition with a common critical coupling (direct transition), so the N\'eel-ordered phase corresponds to the VBS-disordered phase, and conversely, the VBS-ordered phase corresponds to the N\'eel-disordered phase. We now establish that at the transition both order parameters are finite. Figure \ref{mszl-2} depicts the finite-size behavior of the order parameters $M_z^2$ and $\langle \phi^2 \rangle$ close to the phase transition point, up to $L=20$. 
Both parameters converge to finite values at the same coupling ratio $g=0.113$ as the system size tends to infinity, indicating simultaneous transitions. 
This observation leads to the conclusion that a first-order phase transition occurs between two distinct ordered states, the VBS state and the N\'eel state.

\begin{figure}[t]
\centering
\setlength{\abovecaptionskip}{20pt}
\includegraphics[angle=0,width=0.45\textwidth]{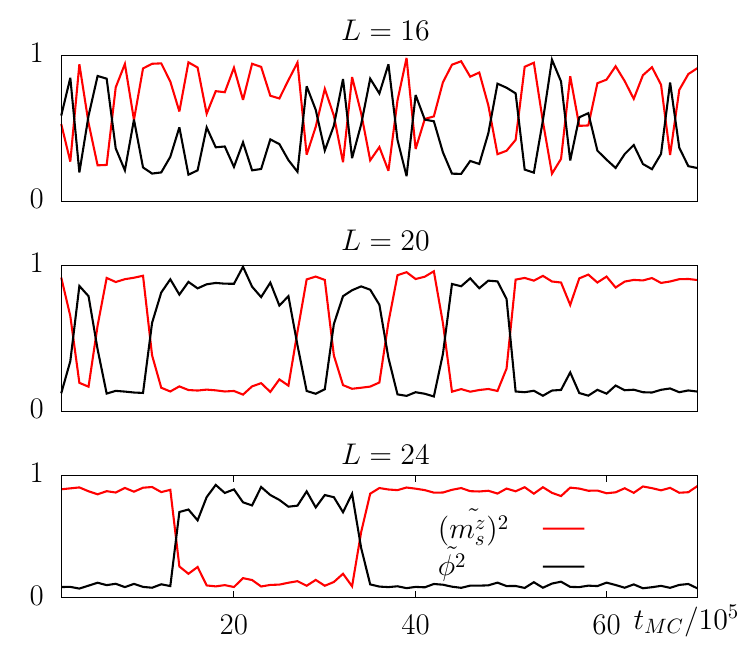}
\vskip 0.4cm
\caption{MC histories of $(m_s^z)^2$ and $\phi^2$ near the phase transition point $g=0.113$ for $L=16, 20, 24$. Each value in the diagram is an average over $10^5$ MC samples. They all show clear switching behavior in both quantities, and the switching time becomes longer as 
system size increases. 
Here $\tilde{(m_s^z)^2}$ and $\tilde{\phi^2}$ are normalized values of $(m_s^z)^2$ and $\phi^2$ such that the maximum is unity. }
\label{mc}
\end{figure}

\subsection{\label{metast}Metastability}

\begin{figure}[t]
\centering
\setlength{\abovecaptionskip}{20pt}
\includegraphics[angle=0,width=0.5\textwidth]{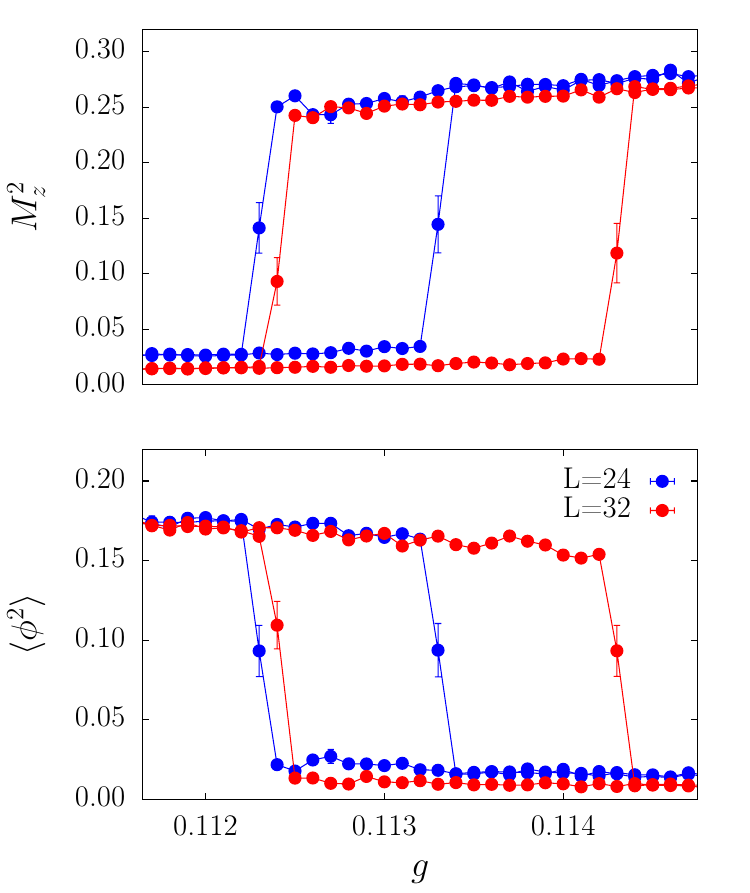}
\caption{The hysteresis loops of the order parameters $M_z^2$ (upper panel)
and $\langle \phi^2 \rangle$ (lower panel). We have used a protocol described in the text where the finite Monte Carlo runs do not explore the phase space ergodically, clearly displaying the phenomena of metastability close to the phase transition, as expected for a first-order transition. }
\label{hy}
\end{figure}

In the preceding subsection on order parameters, we only show data up to $L=20$ due to the considerable difficulty in obtaining reliable statistical averages for $L>20$. This challenge arises from the extended tunneling time between different phases near the phase transition point, which makes ergodic simulations difficult on large lattices.

The long tunneling time can also be studied by simply viewing the time series in the MC binned data. Figure \ref{mc} illustrates the MC histories of $(m_s^z)^2$ and $\phi^2$ with $g=0.113$, near the phase transition point, for system sizes $L=16$, $20$, and $24$. Clear switching behavior is evident in both quantities across all system sizes, with longer switching times observed as the system size increases. Notably, one order parameter predominates when the other is absent, indicating a reciprocal relationship between the ordered states. 
Specifically, the presence of the N\'eel order corresponds to the absence of the VBS order, and vice versa. This phenomenon of switching between the two orders near the phase transition point is characteristic of a first-order transition and suggests a single transition between two distinct ordered states.

The extended tunneling time leads to a hysteresis phenomenon, which we can simulate using the following strategy: initiate simulations at a specific value of $g$ within the VBS ordered phase and incrementally increasing $g$ to transition into the disordered phase, while continuously measuring the order parameters $\langle (m_s^z)^2 \rangle$ and $\langle \phi^2 \rangle$. Subsequently, the final configuration obtained at a particular $g$ serves as the starting point for the subsequent 
simulations incrementally decreasing $g$. 
When the system size $L$ is large, the tunneling time becomes significantly longer compared to the simulation duration. Consequently, the system may persist in a metastable phase even after transitioning beyond the phase boundary. 
Figure \ref{hy} presents the average behavior of the order parameters throughout this process, revealing prominent hysteresis phenomena.

Finally, we note that as shown in Appendix~\ref{JQJmodel} the $\Qth$ term favors the N\'eel order. So, the combination of the $\Qth$ term and $\Qf$ term also realizes the phase transition from VBS to N\'eel. We present the numerical analysis of this model in Appendix~\ref{QJQmodel}. We find that similar to the analysis in this section, the $\Qth-\Qf$ model also hosts a first-order N\'eel-VBS phase transition.

\section{Summary\label{sec:sum}}

In summary, we have studied the N\'eel-VBS transition in sign-free $S=3/2$ models on a square lattice. Utilizing unbiased quantum Monte Carlo numerical methods, we have demonstrated the presence of a direct first-order phase transition from the N\'eel state to the VBS state. 
Our analysis involved studying N\'eel and VBS order parameters, Binder cumulants, dimensionless ratios, and histograms of the two different orders. The abrupt changes in order parameters, the non-analytic and negative behavior of the Binder cumulants, and the mixture of distributions in the histograms near the phase transition point provide compelling evidence of the first-order phase transition behavior.

According to an application of the theory of deconfined criticality, the $S=3/2$ transition should be ``as likely continuous" (first-order transitions can never be ruled out in specific microscopic models even if the transition can be continuous in some models)  as the much discussed $S=1/2$ case, since the effective field theory is identical. Yet we find in our models a strong first-order transition and no evidence even for a weakly first-order transition as has been established in the $S=1/2$ $J$-$Q$ model~\cite{demidiofirst}.

It would be intriguing to find other spin-$3/2$ models in which the N\'eel-VBS phase transition is less strongly first order and look for evidence of scaling, so that the scaling dimensions can be compared with the spin-$1/2$ case~\cite{takahashi2024so5multicriticalitytwodimensionalquantum}. One particular strategy (which can be accessed in the models studied here) would be to begin with the $\Jf-\Qf$ model, which has a continuous or weakly first-order transition~\cite{Lou-SUN,kaulsu34} and then add some $\Jth-\Qth$ terms that reduce the symmetry from SU(4) to SU(2). We leave the exploration of the phase diagram in this parameter regime for future work.\\

\begin{acknowledgments}
We thank A.W. Sandvik for the helpful discussions.
This work was supported by the National Natural Science Foundation of China under Grant No. 12175015 (FZ, WG) and by NSF DMR-2312742 (RKK). This research was supported in part by grant NSF PHY-2309135 to the Kavli Institute for Theoretical Physics (KITP) (RKK).

\end{acknowledgments}


 \bibliography{jq32.bib}

 \clearpage
\appendix
\section{Numerical results on the $\Jth$-$\Qth$ model}
\label{JQJmodel}
Here, we verify that the $\Jth$-$\Qth$ model described by Eq. (\ref{eq:H_JQJ}) does not exhibit a phase transition to VBS order. Employing the same definitions for the order parameters as in Eq.(\ref{eq:phix}), Eq.(\ref{eq:phiy}), and Eq.(\ref{eq:msz2}), we conduct QMC simulations for the spin-3/2 $\Jth$-$\Qth$ model. 

\begin{figure}[htbp]
\centering
\setlength{\abovecaptionskip}{20pt}
\includegraphics[angle=0,width=0.5\textwidth]{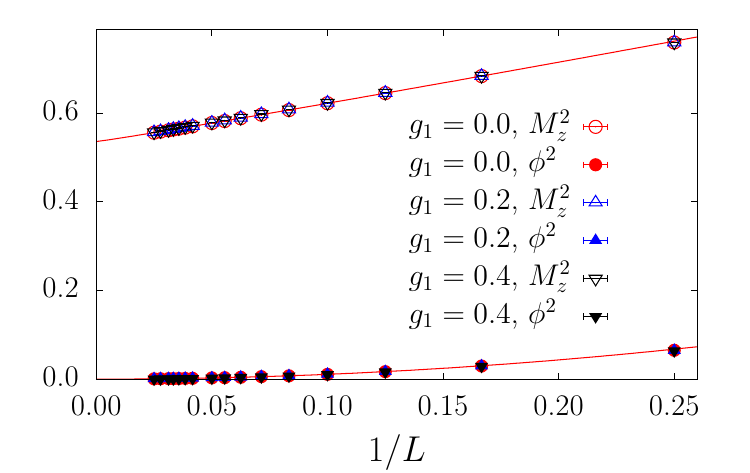}
\vskip 0.4cm
\caption{Finite-size analysis of the order parameters for the spin-3/2 $\Jth$-$\Qth$ model at $g_1=\Jth/\Qth=0, 0.2,0.4$. The value appears to be unaffected by $g$ and tends towards 0 for $M_z^2$, while remaining finite for $\langle \phi^2 \rangle$. The solid line depicted in the figure is fitted using a power-law function $f(L)=a+b/L^c$ at $g_1=0$. This fitting reveals the presence of N\'eel order and the absence of VBS order in the $\Jth$-$\Qth$ model.}
\label{JQJ-lmphi}
\end{figure}

Finite-size analysis of the order parameters for the spin-3/2 $\Jth$-$\Qth$ model at $g_1\equiv\Jth/\Qth=0, 0.2, 0.4$ is depicted in Fig.\ref{JQJ-lmphi}. $g_1=0$ represents the strongest relative $\Qth$ interaction. As observed, $\langle (m_s^z)^2 \rangle$ converges to a finite value, while $\langle \phi^2 \rangle$ tends towards zero as $L$ approaches infinity regardless of the value of $g$. This behavior provides compelling evidence that there is no VBS phase present in the spin-3/2 $\Jth$-$\Qth$ model.

\section{Numerical results on the $\Jf$-$\Qth$ model}
\label{JQQJmodel}

Here we verify that the $\Jf$-$\Qth$ model described does not exhibit a phase transition to a VBS order. We already know that both $\Jf$ and $\Qth$ models are magnetically ordered, so it is unlikely that VBS order can emerge for intermediate couplings, but it is helpful to verify nonetheless. Employing the same definitions for the order parameters as in Eq.(\ref{eq:phix}), Eq.(\ref{eq:phiy}), and Eq.(\ref{eq:msz2}), we conduct QMC simulations for the spin-3/2 $\Jf$-$\Qth$ model. 

\begin{figure}[htbp]
\centering
\setlength{\abovecaptionskip}{20pt}
\includegraphics[angle=0,width=0.5\textwidth]{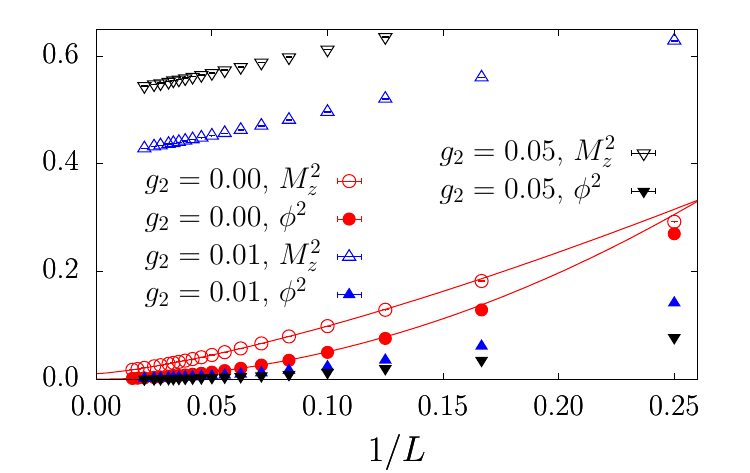}
\vskip 0.4cm
\caption{Finite-size analysis of the order parameters for the spin-3/2 $\Jf$-$\Qth$ model at $g_2=\Qth/\Jf=0, 0.01,0.05$. The solid line depicted in the figure is fitted using a power-law function $f(L)=a+b/L^c$ at $g_2=0$. This fitting reveals the presence of N\'eel order and the absence of VBS order in the $\Jf$-$\Qth$ model.}
\label{JQQJ-lmphi}
\end{figure}

Finite-size analysis of the order parameters for the spin-3/2 $\Jf$-$\Qth$ model at $g_2\equiv\Qth/\Jf=0, 0.01, 0.05$ is depicted in Fig. \ref{JQQJ-lmphi}. When $g_2=0$, the relative $\Jf$ interaction is the strongest. As shown, $M_z^2$ converges to a finite value as $L$ approaches infinity, and this value increases with $g_2$. Conversely, $\langle \phi^2 \rangle$ tends towards zero as $L$ approaches infinity, regardless of the value of $g_2$. This behavior provides compelling evidence that there is no VBS phase present in the spin-3/2 $\Jf$-$\Qth$ model, while the N\'eel order is always present and strengthens with increasing $g_2$.

\section{Numerical results on the $\Qth$-$\Qf$ model}
\label{QJQmodel}
\begin{figure}[htbp]
\centering
\setlength{\abovecaptionskip}{20pt}
\includegraphics[angle=0,width=0.5\textwidth]{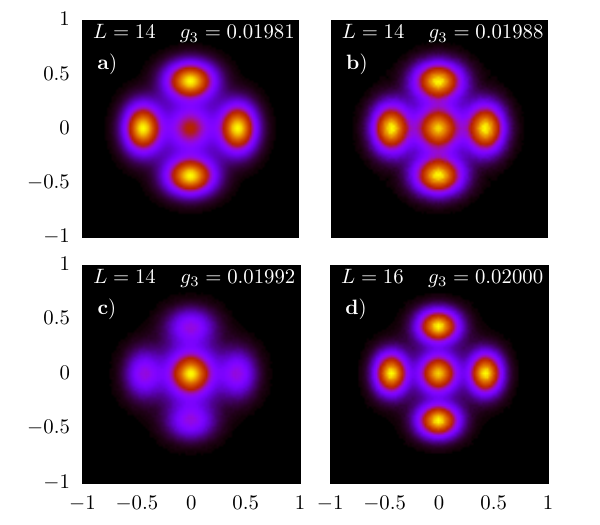}
\caption{The histograms of $\vec\phi$. a), b), and c) for $L=14$ at VBS phase ($g_3=0.01981$), at around phase transition point ($g_3=0.01988$), and at VBS phase ($g_3=0.01992$), respectively; d) for $L=16$, at around phase transition point ($g_3=0.02000$).}
\label{QJQ-hisphi}
\end{figure}

\begin{figure}[htbp]
\centering
\setlength{\abovecaptionskip}{20pt}
\includegraphics[angle=0,width=0.45\textwidth]{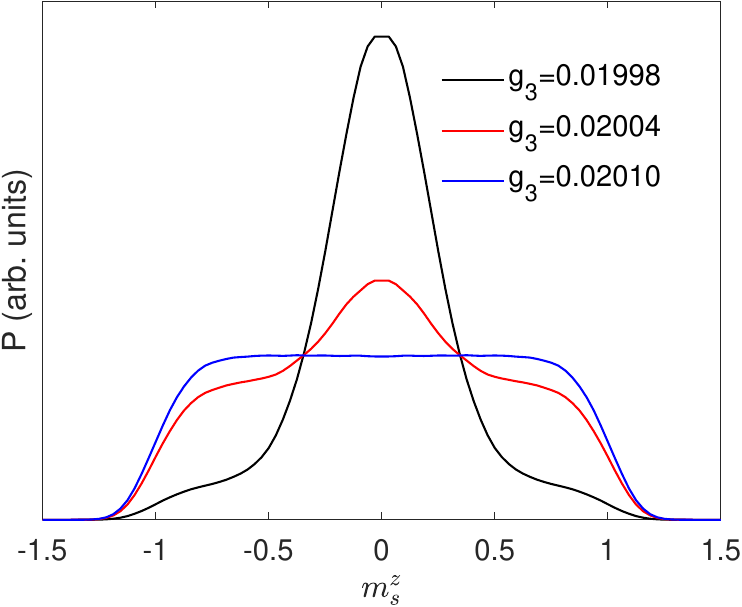}
\vskip 0.4cm
\caption{The histograms of $m_s^z$ for $L=16$ at the N\'eel phase ($g_3=0.02010$), around phase transition point ($g_3=0.02004$), and the VBS phase ($g_3=0.01998$). }
\label{QJQ-hismsz}
\end{figure}

Following the same definitions for the order parameters as in Eq.(\ref{eq:phix}), Eq.(\ref{eq:phiy}), and Eq.(\ref{eq:msz2}), and setting $g_3=\Qth/\Qf$, we also conduct QMC simulations for the $\Qth$-$\Qf$ model. In this model, we are guaranteed a N\'eel-VBS transition because, as we have discussed, the $\Qth$ model is magnetically ordered and the $\Qf$ model is VBS ordered. We only present a brief analysis here since the results are similar to the spin-3/2 $\Jth$-$\Qf$ model presented in the main text, exhibiting a direct first-order phase transition from the VBS ordered phase to the N\'eel ordered phase. The behavior of the order parameters and Binder cumulants is similar to that of the spin-3/2 $\Jth$-$\Qf$ model. To avoid redundancy, we only present the histograms of the two order parameters near the phase transition point in Fig.\ref{QJQ-hisphi} and Fig.\ref{QJQ-hismsz}. 

\end{document}